\documentclass{pasa}%

\title[black hole formation]{Formation of supermassive black hole seeds}
\author[Latif \& Ferrara]{Muhammad A. Latif$^1$\thanks{ latif@iap.fr}, Andrea Ferrara$^2$ \\
\affil{$^1$Sorbonne Universit\'es, UPMC Univ Paris 06 et CNRS, UMR 7095, Institut d'Astrophysique de Paris, 98 bis bd Arago, \\ 75014 Paris, France }%
\affil{$^2$Scuola Normale Superiore, Piazza dei Cavalieri 7, 56126 Pisa, Italy}}%
\jid{PASA}
\doi{10.1017/pas.\the\year.xxx}
\jyear{\the\year}

\usepackage[authoryear]{natbib}
\bibpunct{(}{)}{;}{a}{}{,}
\usepackage{aas_macros}
\usepackage{hyperref} 
\hypersetup{colorlinks,citecolor=blue,linkcolor=blue,urlcolor=blue}

\begin{document}%
\begin{abstract}
The detection of quasars at $z>6$ unveils the presence of supermassive black holes (BHs) of a few billion solar masses.  The rapid formation process of these extreme objects remains a fascinating and open issue.  Such discovery implies that seed black holes must have formed early on, and grown via either rapid accretion or BH/galaxy mergers. In this theoretical review, we discuss in detail various BH seed formation mechanisms and  the physical processes at play during their assembly. We discuss the three most popular BH formation scenarios, involving the (i) core-collapse of massive stars, (ii) dynamical evolution of dense nuclear star clusters, (iii) collapse of a protogalactic metal free gas cloud. This article aims at giving a broad introduction and an overview of the most advanced research in the field. 
\end{abstract}
\begin{keywords}
Black holes -- cosmology -- theory -- quasars -- high redshift universe
\end{keywords}
\maketitle%
\section{INTRODUCTION}
\label{sec:intro}
Observations of quasars at $z > 6$ reveal the existence of supermassive black holes (SMBHs) of a few billion solar masses within the 1 Gyr after the Big Bang \citep{Fan2006,Willott2007,Jiang2008,Jiang2009,MOrtlock2011,Venemans2013,Banados2014,Venemans2015,Jiang2015,Wu2015}\footnote{see \cite{Shankar2016} for potential uncertainties in the measurement of BH masses by a factor of a few. So far, about 50 quasars have been  detected at $z > 6$. SMBHs are also common at the  centres of present day galaxies and may co-evolve with their host galaxies \citep{Kormendy2013,Graham2016}.} How do they form and what are their formation mechanisms are still open questions (see previous reviews on this topic \cite{Volonteri2010,Volonteri2012} and \cite{Haiman2013}).

The presence of SMBHs a few hundred million years after the Big Bang suggests that their seeds must have formed  at $z \geq15$. The masse scale of  BHs depend on the formation mechanism and may vary from $\rm 10 - 10^5~M_{\odot}$.  These seed BHs  must have grown via intense accretion and/or  merging to reach a few billion solar masses.  In order to form a SMBH of  $\rm 2 \times 10^9~M_{\odot}$ at z=7.02,  a seed black should have an initial mass of about 400 $\rm M_{\odot}$ and continuously accrete at the Eddington limit throughout its lifetime. Therefore,  more massive seeds  forming at $z \geq15$ can preferably explain the presence of quasars  at $\rm z \geq 6$.

In this review, we mostly provide a theoretical overview of the field and discuss various astrophysical processes involved in  the formation of seed BHs. We discuss the three most popular BH formation scenarios, involving the (i) core-collapse of massive stars, (ii) dynamical evolution of dense nuclear star clusters, (iii) collapse of a protogalactic metal free gas cloud, also known as the Direct Collapse Black Hole (DCBH) formation channel.  

\section{Black hole formation mechanisms}
Various mechanisms to form black holes have been proposed in the literature since the seminal work by \cite{Rees1984}. They can be classified into three broad categories, which are introduced below and discussed in detail in the following sections. 

The most natural way to form a BH is  the collapse of a  massive star into a BH known as the  ``stellar mass BH". The final mass of BH depends on the metallicity, mass and rotation speed of a star and is ultimately associated with the properties of a native gas cloud \citep{Ciardi2005, Bromm2013}.  BHs may have formed in the dense nuclear clusters  via stellar dynamical processes or relativistic instabilities \citep{Baumgarte1999,PZwart1999,Devecchi2009}. The mass of resulting BH depends on the mass and compactness of a stellar cluster, its dynamical evolution and binary fraction etc.  An alternative scenario could be the monolithic collapse of a protogalactic gas cloud into a massive BH so-called the direct collapse model \citep{Rees1984,Loeb1994,Volonteri2005,Begelman2006,Spaans2006,Latif2011,Ferrara14}. The key requirement for this scenario is to have large gas accretion rates of $\rm \geq 0.1~M_{\odot}/yr$ \citep{Hosokawa2013,Schleicher13}. Such high gas accretion rates can be be achieved in massive primordial halos of $\rm \sim 10^8~M_{\odot}$ illuminated by a strong Lyman Werner flux where in-situ star formation remains suppressed or via dynamical processes  such as a merger of metal rich galaxies or  'bars within bars' instabilities \citep{Begelman2006, Mayer2015}.  

Primordial black holes may have born during the early stages of Big Bang but there is no observational evidence for their existence \citep{Alcock2000,Afshordi2003,Tisserand2007,Ricotti2008}.  In fact, the constraints from microlensing and  spectral distortions of the cosmic micro wave background limit their masses below 1000 $\rm M_{\odot}$.

\section{Stellar mass BHs}
 The first generation of stars so-called population III (Pop III) stars are formed in  minihalos  of $\rm 10^5-10^6~M_{\odot}$ at z $\sim$ 20 - 30. Collapse in these halos is triggered by the molecular hydrogen cooling which brings  the gas temperature down to  about 200 K.  In the absence of dust and metals,  the cooling ability of the primordial gas is considerably reduced and the gas temperature is about a  factor of 10-20 higher compared to the contemporary  star formation in molecular clouds \citep{Abel2002,Bromm2002}.  The thermal Jeans mass scales with $T^{3/2}$ and therefore stellar masses are expected to be higher.  The fraction of molecular hydrogen gets boosted during the collapse by three-body processes and gas cloud becomes fully molecular. In the mean time, gas becomes optically thick to molecular hydrogen cooling and consequently gas temperature rises until the protostar begins to form \citep{Palla1983,Omukai2000}. 
 
 The protostar borns  in  the dense core embedded in molecular hydrogen gas cloud and grows by accretion or even by merging of  dense clumps. In the case of efficient fragmentation, more than one stars are expected to form per halo and we discuss this in detail in the following subsections. Initially, the radius of a protostar increases during the adiabatic accretion phase up to 10 $\rm M_{\odot}$ and then subsequently star enters in the Kelvin-Helmholtz (KH) phase by radiating away its thermal energy.  The interior temperature of star continues to increase until the hydrogen burning starts  around 100 $\rm M_{\odot}$, accretion stops and  star enters  the zero age main sequence (ZAMS) \citep{Stahler1986,OmukaiPalla2001,OmukaiPalla2003,Yoshida2006}.
 
The final mass of a star depends on the properties of natal gas clouds such as its mass, spin, formation redshift and mass accretion rate \citep{Latif2013ApJ,Hirano2014}.  The stars with masses below 9 $\rm M_{\odot}$  do not have enough massive cores to collapse but instead end their lives as white dwarfs.  However, Pop III stars with masses between 25-140 $\rm M_{\odot}$ and above 260 $\rm M_{\odot}$ are expected to directly collapse into a BH \citep{Heger2002,Heger2003}. This is true for the single stars formed out of zero metallicity gas.  However, the presence of metals and rotation may affect their final fates as we discuss below. In the following subsections, we discuss  our current understanding of the initial mass function of Pop III stars, the role of metals and rotation during the  formation and  evolution of stars and their implications for BH formation. Finally, we summarise the expected properties of Pop III remnant BHs.

\subsection{Pop III initial mass function}
\begin{figure*}
\hspace{0.2 cm}
\includegraphics[scale=1.0]{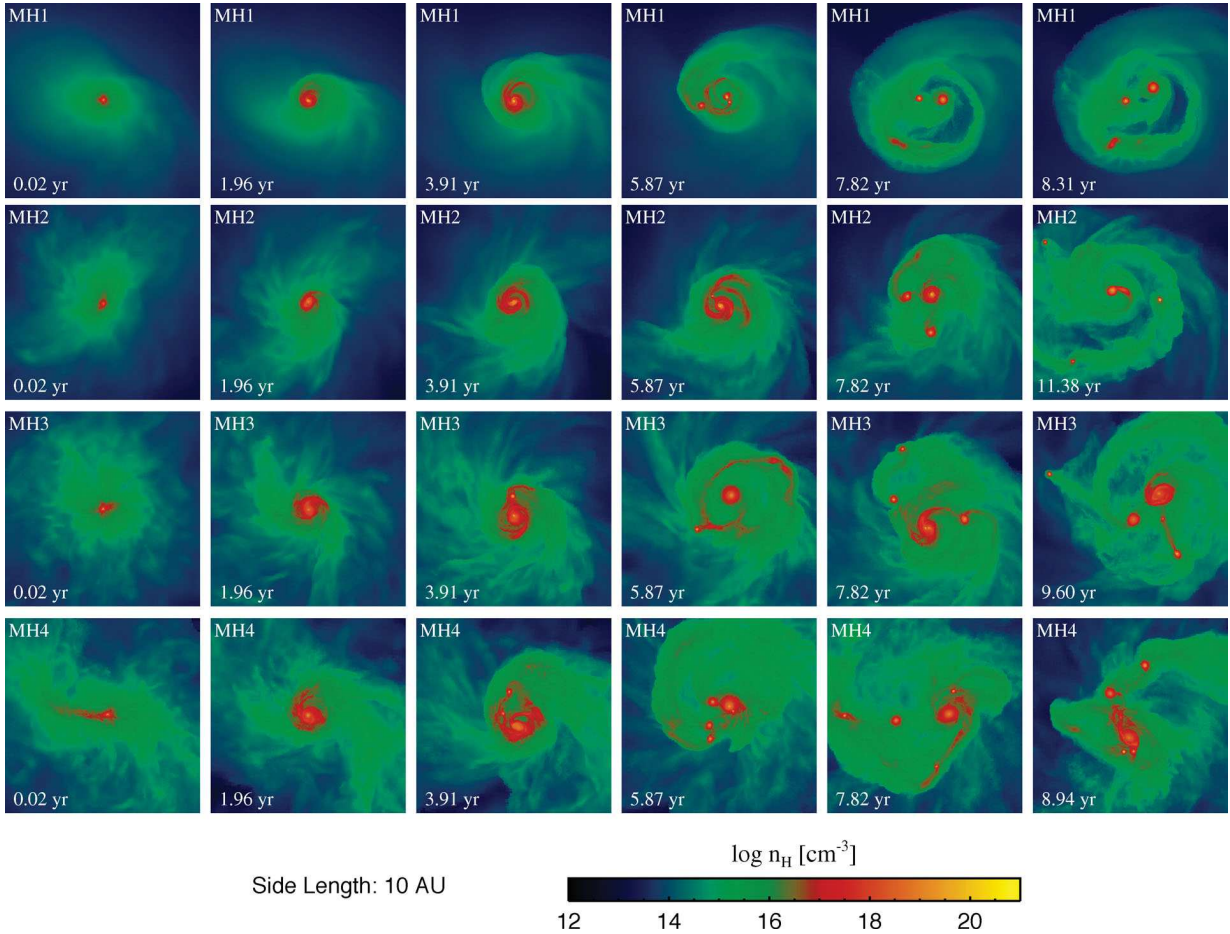} 
\caption{ The evolution of  a protosetllar system  in four different  minihalos.  Density projections of hydrogen nuclei  are shown for  the central 10 AU.  Each row represents the minihalo while each column shows the time evolution after the formation of a central star. Adopted from \cite{Greif12}.}
\label{figure1}
\end{figure*}

Understanding the initial mass function (IMF) is a key challenge in the study of primordial star formation and the masses of stellar BHs are strongly associated with it. The first numerical simulations \citep{Abel2002,Bromm2002} suggested that  primordial stars were massive with typical masses of a  few hundred solar.  The latter studies  also showed  that  Pop III stars were born in isolation with typical mass accretion rates of $\rm 10^{-4}-10^{-2}~M_{\odot}/yr$ \citep{Yoshida2003,Yoshida2006,Oshea2007,Yoshida2008}. However, during the past few years high resolution simulations including  a detailed treatment of physical processes  found that a protostellar disk formed  as a consequence of  gravitational collapse becomes unstable and  fragments into multiple clumps, see Fig. \ref{figure1}. This may lead to the formation of multiple stars per halo \citep{Turk09,Stacy2010,Clark11,Greif12,Latif2013ApJ}. These simulations were evolved only up to a few tens to 5000 years, about a factor of 100 lower than the time required for a protostar to reach the main sequence and also did not include the stellar UV feedback. 

However, simulations taking into account the UV feedback from the protostar show that  it shuts accretion onto the protostar by photoevaporating the protostellar disk and  the central star cannot grow beyond $\rm 40 ~M_{\odot}$ \citep{Hosokawa11,Stacy2012}. These simulations either employed an approximate treatment for the radiative feedback or performed only two dimensional simulations. \cite{Hirano2014} have derived the  stellar mass distribution for one hundred minihalos under the assumption that only single star forms per halo by including the radiative feedback from a star as well as stellar evolution. They found that stellar masses range from 10-1000 $\rm M_{\odot}$ and depend on the properties of their natal halos, also see  \cite{Susa14}.  \cite{Latif2015Disk} employed an analytical model to study the  properties of a protostellar disk around a primordial star and argue that although the disk is susceptible to fragmentation but the clump migration time is shorter than the Kelvin-Helmholtz time scale and therefore clumps may be able to migrate inwards. In fact,  3D radiation hydrodynamical simulations by \cite{Hosokawa2015} show that clump migration leads to intermittent accretion (also see \cite{Vorobyov2010} and \citep{Vorobyov2013}) and disk fragmentation does not halt the formation of massive stars. Therefore, stellar masses may range from a few tens to a few hundred solar.

\begin{figure}
\hspace{-0.2 cm}
\includegraphics[scale=0.4]{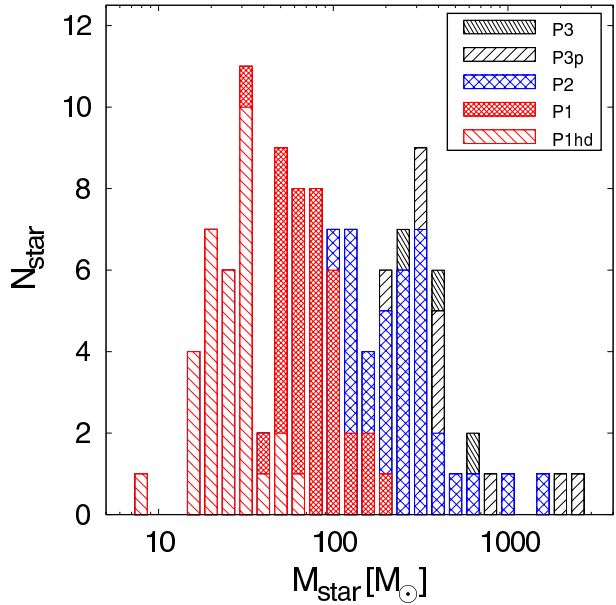} 
\caption{The stellar mass distribution of 110 first stars assuming that single star forms in each minihalo.  Each color represents different stellar evolution path, see \cite{Hirano2014} for details. Adopted from \cite{Hirano2014}. }
\label{figure2}
\end{figure}

The most recent simulations of \cite{Stacy2016} found that disk fragmentation leads to the formation of a stellar cluster with a top heavy IMF. In their simulations, the most massive star reaches 20 solar masses in 5000 yrs after its formation and some of the sinks get ejected from the disk before ionisation front breaks out. However, these simulations were evolved only for 5000 yrs after the formation of the first sink and therefore final masses of Pop III stars are still uncertain.  In Fig. \ref{figure2}, we show the stellar mass distribution from \cite{Hirano2014} which gives an upper limit on the expected stellar masses as their calculations do not take into account the multiplicity of stars per halo. These results suggest that  the typical mass of Pop III stars is about 100 $\rm M_{\odot}$ with the exception of a few cases of  a 1000 $\rm M_{\odot}$. 
%Therefore,  a hundred solar mass sized BHs are expected to form in this scenario.

\subsection{Effects of metallicity and rotation}
Both  the metallicity and the rotation speed of stars play a vital role in defining their fates and  have important implications for the formation of stellar mass BHs. In the presence of trace amounts of metals, dust cooling becomes important at high densities, and triggers the formation of multiple low mass stars \citep{Ferrara2000,Schneider2003,Omukai2005,Schneider2006}. Numerical simulations  show that  for $\rm Z/Z_{\odot} \geq 10^{-5}$ dust cooling  induces fragmentation  and  fosters low mass star formation \citep{Dopcke2011,Dopcke2013,Smith2015}.  Similarly, the  dark matter halos with higher spin have a longer collapse timescale which results in enhanced fragmentation. Moreover,  higher rotation decreases the mass accretion onto a protostar and consequently the final stellar mass gets reduced \citep{Hirano2014,Dutta2016}.

The presence of metals and rotation does not only  influence the formation of the first/second generation of stars by reducing their final masses but also strongly affects their evolution. Stellar evolution models show that  mass loss from the stellar winds is metallicity dependent, scales with $\rm \dot{m}_{loss} \propto Z^{0.5}$ and consequently metal rich stars show much higher mass loss compared to the metal poor stars \citep{Nugis2000,Kudritzki2000,Baraffe2001,Heger2003,Meynet2005}. Similarly, fast rotation enhances the surface enrichment of CNO cycle elements which in turn derive the mass loss  by stellar winds from metal poor stars. For example, a fast rotating 60 $\rm M_{\odot}$ star can lose 30-55\% of its initial mass for  $Z/Z_{\odot}=10^{-8}-10^{-5}$ \citep{Heger2000,Meynet2000,Meynet2006,Meynet2006A,Chiappini2011}. Furthermore, an enhanced rotation modifies the Eddington limit known as the $\Omega \Gamma$ limit \citep{Langer1997,Meynet2000} and prevents the star from growing beyond $\rm 20- 40 ~M_{\odot}$ by making it more compact \citep{Lee2016}.  Such rapidly rotating low metallicity stars  may  also produce  gamma-ray bursts \citep{Yoon2005}.

\subsection{Stellar  tracks leading to BH formation}
Stellar evolution calculations suggest that  the growth of  primordial stars is significantly different from the present-day counterparts. Larger mass accretion rates, about two-three orders of magnitude higher than normal stars, and the absence of heavier elements makes  the evolution of Pop III stars significantly different from ordinary stars \citep{OmukaiPalla2001,Schaerer2002,OmukaiPalla2003,Schaerer2003}.  Stellar evolution also strongly depends on the time evolution of  the mass accretion rates. Deuterium burning and pp cycle do not  generate enough energy to counteract the KH contraction.  The hydrogen burning from the CN cycle significantly increases the stellar luminosity and a protostar quickly reaches the ZAMS. Therefore, more massive stars of the order of a  hundred solar masses are expected to form from a zero metallicity gas while for $\rm Z/Z_{\odot} \geq 10^{-2}$  radiation pressure on dust grains becomes important and leads to the formation of low mass stars \citep{OmukaiPalla2003,Hosokawa2012}. 

\begin{figure}
\hspace{-0.2 cm}
\includegraphics[scale=0.6]{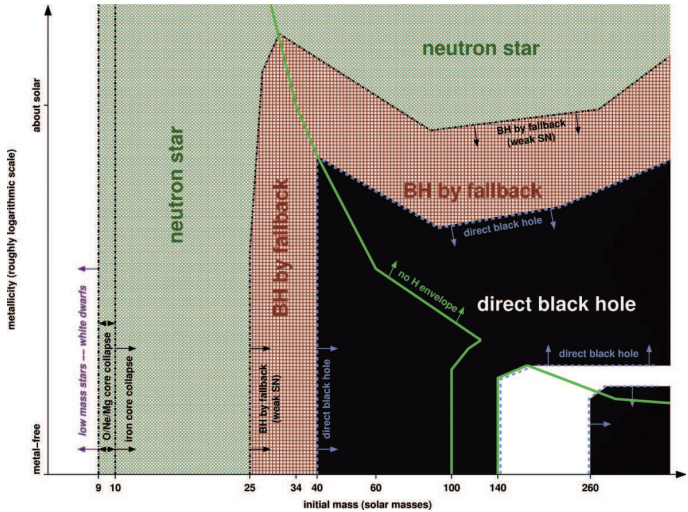} 
\caption{ The fate of single stars as a function of their initial mass and initial metallicity.  The tracks for the formation of direct BHs from the stars are highlighted by the black color while the white region in the bottom right indicates the range for a pair instability supernova. Adopted from \cite{Heger2003}. }
\label{figure21}
\end{figure}

The mass loss from stars become significant even in the presence of trace amount of metals while for primordial stars such losses are minimal. Stellar evolutionary tracks leading to the formation of BHs are indicated in Fig. \ref{figure21}. The primordial stars with masses between 40-140 $\rm M_{\odot}$ and above 260 $\rm M_{\odot}$ are expected to directly collapse into BHs of similar masses \citep{Heger2002,Heger2003}.  For a trace amount of metals all stars above 40 $\rm M_{\odot}$ directly collapse in a BH but with increasing metallicity the mass loss starts to increase. Moreover, the fraction of massive stars collapsing into a BH depends on the shape of the IMF and is about twice for top-heavy IMF compared to the Salpeter IMF \citep{Heger2003}.  Although numerical simulations suggest that  some of the Pop III stars might be rotating at $\rm \geq 1000~ km/s$, i.e.  close to their break up limit \citep{StacyRot2011}, stellar evolution calculations including rotation \citep{Ekstrom2008}  indicate that the mass loss  from the fast rotating Pop III stars is still very low. In a nutshell, metal free stars with low rotation speeds are favoured for the formation of stellar mass BHs as they retain most of their mass until they collapse into a BH.

\subsection{Expected properties of stellar mass BHs}
 The current numerical simulations indicate that the formation of massive primordial stars up to a 1000 $\rm M_{\odot}$ is possible with the characteristic mass scale between 10-100 $\rm M_{\odot}$. This suggests that  BH seeds of  a few  hundred solar masses can be formed at $z=20-30$.  In order to reach a few billion solar masses by $z=7$, they must  continuously grow at the Eddington  limit. The three dimensional numerical simulations studying the growth of stellar mass BHs show that feedback from  a BH photo-evaporates the gas the in its hosting halo and consequently accretion onto a BH gets  halted \citep{Johnson2007,Alvarez2009}.  They further found that  the BH accretion rate is about  $\rm 10^{-10}~M_{\odot}/yr$, i.e. several orders of magnitude below the Eddington limit which makes their growth extremely difficult. Moreover, the progenitor star creates an HII region, evacuates the gas from a minihalo and accretion remains halted for about $\rm 10^8$ yrs.  \cite{Park2016} found that stellar mass black holes cannot coevally grow  with budge via accretion until it reaches the critical mass of $\rm 10^6~M_{\odot}$. There is a growing consensus that  the stellar mass BHs require various episodes of super Eddington accretion to reach a billion solar masses  by $z \geq 6$ \citep{Madau2014,Volonteri2015,Pacucci2015,Inayoshi2016}.

\section{BH seeds from dense stellar clusters}
The self-gravitating stellar systems with negative heat capacity  are susceptible to a gravitational collapse where the core collapse time scale ($\rm t_{cc}$) is comparable to the two-body relaxation time \citep{Binney1987}. In such a scenario the core of the cluster collapses resulting in a higher  stellar density at its centre and the run-away stellar collisions lead to the formation of a very massive star (VMS) which later may collapse into a massive seed BH of up to a 1000 $\rm M_{\odot}$. This mechanism provides an additional route for the formation of massive BH seeds. The earlier studies suggested that  dense stellar clusters with  $\rm \geq 10^7$ stars are required for the run-away collapse to occur and subsequent growth of a massive object \citep{Lee1987,Quinlan1990}. It was found that the binary heating stops the core collapse in less massive stellar clusters \citep{Heggie1975,Hut1992}. Moreover, the $\rm t_{cc}$ has to be shorter than the typical time scale for the massive stars to reach the main sequence ($\rm \sim$ 3 Myrs) otherwise mass loss from the supernova may  halt the core collapse.

However, later studies found that the mass segregation instability can lead to the formation of a massive black hole seed even in less massive clusters \citep{Spitzer1969,Vishniac1978,Begelman1978}. This comes from the fact that massive stars sink into the centre of a cluster on a dynamical friction time scale and speed up the collapse. This phenomenon is expected to get accelerated significantly  in multi-mass systems like realistic nuclear clusters. If the mass segregation occurs within  the first three Myrs, then stellar dynamical processes can lead to the formation of a VMS. As we discuss in the following sections, this mechanism strongly depends on the compactness and dynamical evolution of a stellar cluster and is particularly expected to occur in low metallicity clusters where mass loss from stellar winds is expected to be minimum \citep{Hirschi2007,Glebbeek2009}. These dense nuclear clusters can form in   halos of about $\rm 10^8~M_{\odot}$  at $z\sim 15$  enriched by trace amount of metals \citep{Omukai2008,Devecchi2009}. 
 
Another potential way could be the merging of many stellar mass BHs  or a supra-exponential growth of stellar mass black holes in a  cluster by dense cold gas flows \cite{Alexander2014}. However, the gravitational recoil velocities are about  the order of $\rm \sim 1000 ~ km/s$  which are much larger than the escape velocity of halos (10 km/s) and therefore lead to the ejection of BHs from the shallow DM potentials at earlier comic times \citep{Haiman2004,Tanaka2009ApJ}. Alternatives could be the run-away merging of stellar mass black holes  in the core of a dense cluster mediated by the extremely large inflows of gas \citep{Davies2011,Lupi2014}  or  collapse of a dense stellar cluster due to the relativistic instability \citep{Shapiro1986}.

\subsection{Binarity  of  early star formation}
 Observations of  contemporary star formation suggest  that about 50\% of the stars are born in binaries or multiple systems \citep{Sana2009,Sana2011}. In the context of primordial stars, recent numerical simulations show that the first generation of  stars may also have formed in multiple systems  due to the fragmentation of  protostellar disk \citep{Turk09,Stacy2010,Clark11,Greif12,Latif2013ApJ,Latif2015Disk,Stacy2016}.  Some of the clumps from disk fragmentation migrate inward and merge with the central protostar while the rest survive to form a multiple system \citep{Greif12,Stacy2016}.  In fact, numerical simulations suggest that  up to 35\% of Pop III stars may have formed in binaries; thus the probability for a PopIII star to have a companion is about 50\% \citep{Stacy2010,Stacy2013}. This may also suggest the formation of X-ray binaries at earlier cosmic times \citep{Mirabel2011,Ryu2016}.

Simulations of the first galaxies show that supernovae from Pop III stars quickly  enrich halos  with metals where  gas can be cooled down to the  cosmic microwave background  temperature ($\rm T_{CMB}$) by dust and metal line cooling  at z=15 even in the presence of UV flux \citep{Tornatore2007,Smith2009,Greif2010,Wise2012,Whalen2013b,Johnson2013,Safranek-Shrader2014,Bovino14,Pallottini2014}. Stellar clusters of Pop II stars are expected to form above the critical value of the metallicity,  i.e. $\rm Z/Z_{\odot} \sim 5 \times10^{-4}$\citep{Schneider2003,Omukai2005,Cazaux2009,Latif2012}. The recent 3D cosmological simulations suggest that the first bona fide stellar clusters form at $z=15$ where cooling is  triggered by the metal lines (such as CII and OI) and stellar masses range from $\rm 0.1-10~M_{\odot}$ \citep{Safranek-Shrader2014b,Safranek-Shrader2016}. Although, current cosmological simulations are unable to constrain the binary fraction in the first stellar clusters due to the numerical constraints but the  binary fraction is expected to be  higher for metal poor stars \citep{Komiya2007,Machida2008}.

\subsection{Compact stellar clusters}
The compactness of a stellar cluster is the key property for the stellar dynamical processes to occur as it determines whether  a seed BH can form or not.  This requirement arises from the fact that the time for the core collapse should be shorter than the time for massive stars to go off supernova ($\sim 3 ~Myr$) otherwise latter may halt the core collapse, also see \cite{Yajima2016}. The more compact the cluster shorter the core collapse time scale.  For a given cluster mass, the number of collisions  and the increase in mass per collision strongly depend on the half-mass radius of a cluster \citep{Zwart2002}.  The  mass of a VMS almost linearly increases with the initial central density of  a cluster \citep{Katz2015}. 

In the context of  BHs, the first compact nuclear cluster are expected to form in  massive halos of  $\rm  10^8~M_{\odot}$  at $z > 10$ with  $\rm Z/Z_{\odot} \sim10^{-5}-10^{-3}$. In the presence of a  Lyman Werner flux, the formation of $\rm H_2$ at low densities remains suppressed and gas initially collapses isothermally  to form a self-gravitating accretion disk at the centre of a halo \citep{Devecchi2009}. Fragmentation occurs only in the nucleus of the disk above densities of $\rm 10^3~cm^{-3}$ due to the metal line cooling and forms a  compact  cluster of $\rm 10^5 ~M_{\odot}$  with half-mass radius of about 1 pc.  On the other hand for higher metallicities fragmentation already occurs at densities $\rm < 10^3~cm^{-3}$ and results in the core collapse time longer than 3 Myrs \citep{Omukai2008,Devecchi2009,Devecchi2012}.  

Semi-analytical models suggest that large accretion rates and efficient star formation are required  for the formation of a  such compact nuclear stellar and conditions for their formation are feasible in the first massive halos polluted by a trace amount of metals  at $z>10$ \citep{Devecchi2009}. \cite{Latif2016dust} have performed 3D cosmological simulations  to study the impact of trace amount of dust and metals in massive halos of $\rm 10^{8}~M_{\odot}$  and  found that a dense cluster may also  form for $\rm Z/Z_{\odot} \geq 10^{-4}$ in the presence of a strong UV flux.  The study of first nuclear clusters in still in infancy as their formation cannot be resolved in numerical simulations due to the  enormous range of spatial scales. In future, more work is required to asses how compact and massive the clusters can be formed at high redshift.

\subsection{ Dynamical evolution}
The dynamical evolution  of dense stellar clusters  has been studied via direct N-body simulations which show  that  the mergers between stars destroy  binaries and avoid three body binary heating \citep{PZwart1999}. Moreover, in contrary to the previous studies, stellar collisions are fostered by the dynamically formed binaries which increase the star collision rate \citep{PZwart1999,Zwart2002}. Consequently, stellar collisions can occur even in a low mass system with 12,000 stars and may lead to the formation of a BH of  up to $\rm \sim 1000~M_{\odot}$.  This was further confirmed from  the Monte Carlo simulations by \cite{Gurkan2004}  where they explored  a range of IMFs and cluster structural parameters. Their followed up work including stellar dynamics  showed that core collapse occurred on  $\rm  \sim 3~ Myr$ time scale irrespective of the cluster size, mass, concentration and  a VMS reached up to $\rm 1000~ M_{\odot}$ \citep{Freitag2006}. Moreover, the mass of a VMS is about $\rm 10^{-3}$  times the cluster mass and is mainly contributed by the stars in the mass range  of 60-120 $\rm M_{\odot}$ \citep{Goswami2012}. 

Metallicity also plays an important role in the dynamical evolution of a stellar cluster as mass loss from stars can deeply affect the core collapse. \cite{Schulman2012} and \cite{Downing2012} found that the size of cluster depends on the metallicity and metal rich clusters expand more rapidly. For more massive clusters, \cite{Sippel2012} found that  both metal poor and metal rich clusters are structurally similar and  no significant differences in their half-mass radii were observed. The effect of metallicity seems to be more  important in intermediate mass young clusters and leads to the differences in the core radius as well as in the half-mass radius \citep{Mapelli2013}. These differences arise from an interplay between the mass loss from stellar winds, dynamical heating and three-body interactions. Therefore, low metallicity young clusters forming at $z>10$ are preferred for the formation of a massive seed  black hole.

\subsection{Expected properties of seed BHs}
The BH mass resulting from the core collapse of a dense stellar cluster depends on its initial stellar mass, initial stellar density, compactness, cluster geometry, fraction of primordial binaries and the IMF of a cluster. The N-body simulations exploring the range of above mentioned parameters show that  clusters with different geometries, fractal distributions and density profiles converge within 1 Myr and do not significantly influence the mass of a VMS. However, the final mass of a VMS  strongly depends on the compactness for a given cluster mass and almost linearly increases with initial central density. Recently, \cite{Katz2015} used a  combination of hydrodynamical  cosmological simulations  and direct N-body simulations  to estimate the mass of  a VMS forming via stellar run-away collisions in a metal poor stellar cluster. They found that  a VMS of $\rm \geq 400 ~ M_{\odot}$ can be formed in  a stellar cluster of  $\rm 10^4 ~M_{\odot}$  at $z=15$ and may reach  up to a 1000 $\rm M_{\odot}$ for a cluster mass of a few times $\rm 10^5~M_{\odot}$. 

The IMF of a stellar cluster  influences the  number of collisions between stars and the mass of a VMS tends to increase  for a top heavy IMF  but the probability of producing a VMS remains the same. Similarly, the introduction of primordial binaries and the initial mass segregation significantly change the evolution of a cluster but the mass of a VMS differs only by a factor of 3 which is less significant than the changes in the central density and the mass of a cluster \citep{Katz2015}. The  mass of a seed black hole forming from the core collapse of a cluster can be estimated  as follows \citep{Zwart2002},
\begin{equation}
 M_{BH} = m_{*} + 4 \times 10^{-3} f_{c} M_{c0} \gamma \ln \Lambda_C ~.
\label{eq}
\end{equation}
Here $m_{*}$ is the mass of a massive star in the cluster, $f_c$ is the  fraction of dynamically formed binaries,  $\rm M_{c0}$ is  the birth mass of  the cluster, $\ln \Lambda_c$ is the Coulomb logarithm, and $\gamma  \sim 1$, is the ratio of time scales, see \cite{Zwart2002} for details. For a $\rm M_{c0}= 10^4 ~M_{\odot}$, $m_{*} =100~M_{\odot}$,  $ f_c =0.2$, $ \ln \Lambda_C =10$,  $\rm M_{BH} \sim 180 ~M_{\odot}$ while for $\rm M_{c0} = 10^5 ~M_{\odot}$, the expected black hole mass  is  $\rm \sim 900 ~M_{\odot}$.

\begin{figure}
\hspace{-0.2 cm}
\includegraphics[scale=0.4]{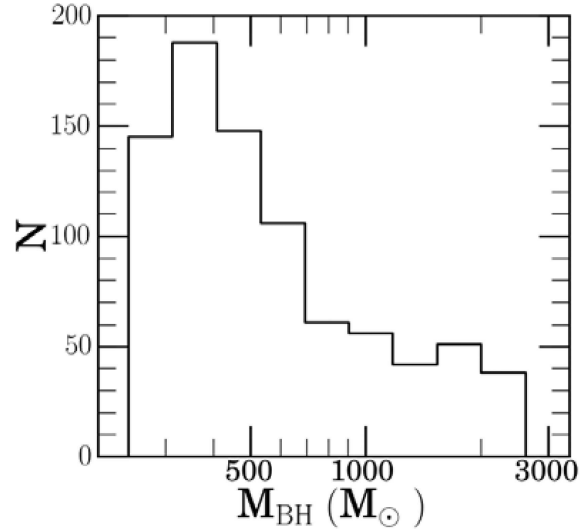} 
\caption{ The mass function of BHs formed via stellar dynamical process in the first nuclear cluster at $z \sim 15$. Adopted from \cite{Devecchi2012}.}
\label{figure3}
\end{figure}

 \cite{Devecchi2010,Devecchi2012} employed an analytical model to estimate the expected mass range of seed BHs forming via dynamical processes in  nuclear stellar clusters  at $z \geq 10$. Their estimates for BH masses are shown in Fig. \ref{figure3} and indicate that BHs of up to a 1000 $\rm M_{\odot}$ can be formed. Simulations self-consistently modelling  the formation and evolution of a nuclear stellar cluster are necessary to better understand the dynamics of the first dense nuclear clusters.

\section{Direct collapse BHs}
One of the most promising way to explain the existence of $z>6$ quasars is to form a massive BH seed of $\rm 10^5-10^6~M_{\odot}$ directly via the gas dynamical processes  \citep{Rees1984,Haehnelt1993,Loeb1994,Eisenstein1995,Bromm03,Koushiappas2004,Begelman2006,Spaans2006}  known as  the direct collapse black hole (DCBH).  The key requirement for this scenario to work is that gas should  efficiently shed angular momentum and rapidly collapse avoiding fragmentation. The basic idea behind this mechanism is to bring  large inflows of gas to the centre of  the halo  on a short time scale  of the order of 1 Myr and let the huge reservoir of the gas to collapse into a single massive object without fragmenting into stars. The recent calculations suggest that large mass accretion rates of $\rm \geq 0.1~ M_{\odot}/yr$ (see our explanation below) are required for this mechanism to work \citep{Begelman2010,Ball2011,Hosokawa2013,Schleicher13,Ferrara14,Sakurai2015}. Such large accretion rates can be obtained either via thermodynamical processes by keeping the gas warm  as the accretion rate $\propto  T^{3/2}$, \citep{Bromm03,Volonteri2008,Regan09,Shang2010,Johnson2011,Latif2013c,Ferrara14}, or through highly dynamical processes such as the 'bars within bars' instability \citep{Shlosman1989,Begelman2006} and galaxy mergers  \citep{Mayer2010}.

The thermodynamical way to form a DCBH requires that halos should be metal free otherwise trace amount of metals  can cool the gas which later leads to fragmentation and star formation. Even in  primordial halos the formation of molecular hydrogen should remain suppressed to keep the gas  warm  and cooling should mainly proceed via atomic lines. Alternatives could be the trapping of Lyman alpha photons which could stiffen the equation of state and suppress in-situ star formation \citep{Spaans2006,Latif2011}. However,  it has been found that cooling can still proceed via the 2s-1s transition and compensates the effect of Lyman alpha trapping \citep{Schleicher10,Latif2011b}. \cite{Lodato2006} suggested that a few percent of pre-galactic disks forming in the dark matter halos at high redshift may form a massive object depending on the properties of hosting halos such as spin, mass and cooling properties. Particularly, low spin halos are more prone to forming a massive central object. \cite{Tanaka2013,Tanaka2014} have proposed that extreme streaming velocities may facilitate direct collapse by enhancing the critical mass of halos to collapse, suppressing in-situ star formation and consequently  avoid metal enrichment in atomic cooling halos. They argue that this scenario  may not need external UV flux to suppress star formation, but \cite{Latif2014Stream} show that the impact of streaming is not significant in atomic cooling halos forming at $z =15$. \cite{Inayoshi2015b} propose that high velocity collisions of two protogalaxies shock heat the gas which later cools isobarically, molecular hydrogen gets collisional dissociated \citep{Inayoshi2012} and  consequently an isothermal collapse may form a supermassive star.  However, \cite{Visbal2014b} and \cite{Fernandez2014} show that these scenarios require additional mechanism to suppress the formation of molecular at high densities.

\cite{Begelman2009} argue that  fragmentation can also be suppressed in  the presence of supersonic turbulence  even in metal rich halos and may not require above mentioned conditions. \cite{Mayer2010} propose that merging of two metal rich galaxies  brings large accretion rates of $\rm \sim 10^4~M_{\odot}/yr$  driven by the  gravitational torques and a compact stable nuclear disk forms which later  may collapse into a massive BH.  \cite{Ferrara13} argue based on one-dimensional model that such disk rapidly cools and becomes unstable in about 100 yrs. So,  the central core cannot grow beyond 100 solar masses.  In the recent study \cite{Mayer2015} improved on their previous work by employing radiative cooling for both optically thin and thick regimes instead of effective equation of state as in  \cite{Mayer2010}. Their new findings suggest that  inclusion of cooling fosters the formation of compact nuclear disk and central core is stabilised by the shock heating and high optical depth of the gas. It is  expected that in such scenario the  core may directly collapse via  general relativistic instabilities into a massive black hole. \cite{Bonoli2014} employed galaxy formation model in Millennium  simulations using the recipe of \cite{Mayer2010} that  major mergers of gas rich  and disk dominated galaxies form massive  BH.  They found that most of the $\rm > 10^{11}~M_{\odot}$ halos at z $\sim$ 4 meet this criteria.
\cite{LatifVolonteri15} and \cite{Latif2015Disk2} show that complete isothermal  monolithic collapse  of a protogalactic gas may always not be necessary to form a massive  central BH.

 The  gas has to shed angular momentum to collapse to high enough densities  to form a massive BH otherwise collapse  gets halted by the angular momentum barrier. \cite{Eisenstein1995} and \cite{Koushiappas2004} proposed that BHs can either form in DM halos with low angular momentum or in the tail of low angular momentum gas within the halo. However, even then gas has to transport angular momentum quite efficiently to form a massive central object. The recent simulations suggest that the triaxility of DM halos exerts gravitational torques which helps in the transfer of angular momentum via 'bars within bars' instabilities \citep{Choi2013,Choi2015}. In the following subsections, we discuss physical processes involved in the formation of DCBHs via isothermal collapse, how and under what conditions they are formed and what are their typical masses.

\subsection{Birthplaces of DCBHs}
 The potential embryos for the formation of DCBHs via isothermal collapse are the first massive metal-free halos with $\rm T_{vir} \geq 10^4$ K and masses of $\rm > 10^7 ~M_{\odot}$  at  $z \sim 15$. They have sufficient gas reservoir to feed a massive object  and  their potential wells are deep enough to foster  a rapid collapse required for the formation of a DCBH. Moreover, their virial temperature is high enough for  the  atomic line cooling to operate. The formation of a DCBH mandates that they should be of a primordial composition and the formation of $\rm H_2$  remains suppressed  \citep{Shang2010,Petri2012,Latif2014UV,Yue2014}.  In the absence of $\rm H_2$ cooling, collapse proceeds isothermally with $T  \sim$ 8000 K and warm gas flows toward the centre at the rate of $\rm \dot{m} \sim {c_s^3}/{G} \sim 0.1~M_{\odot}/yr \left( {T}/{8000~K}\right)^{3/2}$,  where $c_s$ is the thermal sound speed.
 
The suppression of molecular hydrogen requires the presence of a strong Lyman Werner (LW) flux \citep{Omukai2001,Omukai2008,Shang2010,Latif2013}.  As we discuss in the next subsections that constraint of a strong LW fluxes requires that DCBH hosting halo should form in the vicinity (about few kpc) of a massive star forming galaxy \citep{Dijkstra2008,Agarwal2012,Habouzit2016a}. In mean time, it has also to avoid the metal pollution for the above mentioned reasons.  Under these conditions, an isothermal monolithic collapse is expected to occur which later may lead to the formation of a DCBH.

\subsection{UV flux constraints}
One of the main constraints for the formation of DCBHs via isothermal direct collapse is the suppression of molecular hydrogen formation which requires the presence of  a strong LW flux \citep{Omukai2001,Omukai2008,Shang2010,Petri2012,Latif2013}.  The gas phase reactions in primordial gas can lead to the formation of a trace amount of $\rm H_{2}$. The  main pathway for $\rm H_{2}$ formation is:
\begin{equation}
%\mathrm{ H + e^{-} \rightarrow H^{-} + \gamma} \\
\mathrm{H + e^{-} \rightarrow H^{-} +} \gamma
\end{equation}
\begin{equation}
\mathrm{ H + H^{-} \rightarrow  H_{2} + e^-.}\\ 
\label{h21}
\end{equation}
The formation of $\rm H_{2}$ can be suppressed either by directly dissociating  $\rm H_{2}$ or indirectly via the photo-detachment of $\rm H^{-}$. The destruction channel for $\rm H_2$ depends on the stellar spectra. The photons with energy between 11.2-13.6 eV can be absorbed in the Lyman-Werner bands of $\rm H_{2}$ and photo-dissociate it shortly after putting it into an excited state, this is known as the Solomon process. While the low energy photons with energy above $0.76$ eV can photo-detach $\rm H^{-}$.  The reactions for the both processes are the following:
\begin{equation}
%\mathrm{ H + e^{-} \rightarrow H^{-} + \gamma} \\
\mathrm{H_{2}} + \gamma_{LW} \mathrm{\rightarrow H + H}
\label{h20}
\end{equation}
\begin{equation}
\mathrm{H^{-}} + \gamma_{0.76} \mathrm{\rightarrow H + e^{-}}\\ 
\label{h2}
\end{equation}
The  stars with hard spectrum of  $\rm T_{rad}=10^5$ K are more efficient for the direct dissociation of $\rm H_2$ while the  stars characterised with $T_{\rm rad}=10^4$ K are more effective in the photo-detachment of $\rm H^-$. The complete quenching of  $\rm H_2$ requires a critical value of UV flux  ($J_{21}^{\rm crit}$) which depends on the shape of a radiation spectrum. The previous studies used idealised spectra to compute the $J_{21}^{\rm crit}$ and found that $J_{21}^{\rm crit} = 30-1200$ for  $T_{\rm rad}=10^4$ K and $ J_{21}^{\rm crit} = 1000 $ for $T_{\rm rad}=10^5$ K \citep{Omukai2001,Shang2010,VanBorm2013,Latif2014UV,Johnson2014}.  It has been found that  the strength of $J_{21}^{\rm crit}$ is about an order of magnitude above the background UV flux and can only be achieved in the close vicinity of a star forming galaxy \citep{Dijkstra2008,Agarwal2012,Habouzit2016a}. Moreover, such flux is mainly provided by  Pop II stars due to their long lives and higher abundance. Recently,  realistic spectra of the first galaxies was computed using  the stellar synthesis code STARBURST \citep{Leitherer1999} and it was found that $J_{21}^{\rm crit}$ depends on the mode of a star formation (either bursty or constant) as well as on the age and metallicity of stars \citep{Sugimura14,Agarwal2015,Agarwal2015B}. \cite{Sugimura14} show that such spectra can be mimicked with $T_{\rm rad} {\rm = 2 \times 10^4-10^5}$ K.

\begin{figure}
\hspace{-0.4 cm}
\includegraphics[scale=0.5]{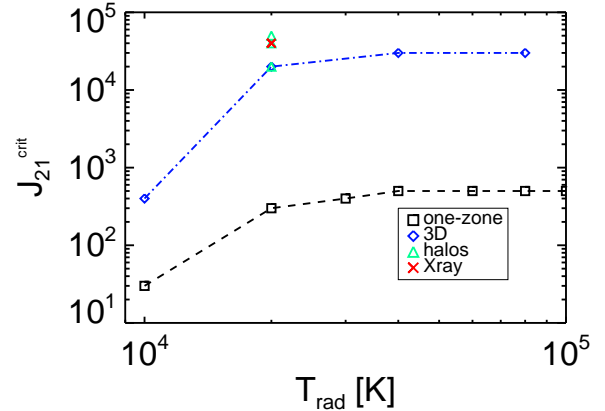} 
\caption{The estimates of critical value of UV flux ($J_{21}^{\rm crit}$)  both from one zone models and 3D simulations including variations from halo to halo, dependence on  the radiation spectra and the impact of X-ray ionization. Adopted from \cite{Latif2015a}.  }
\label{figure4}
\end{figure}

Recent estimates of $J_{21}^{\rm crit}$ from 3D cosmological simulations \citep{Latif2015a}\footnote{The details of chemical network and reaction rates are given in \cite{Latif2015a}, see \cite{Glover2015a,Glover2015b} for their impact on  $J_{21}^{\rm crit}$ and reduced chemical network.} for a realistic Pop II spectra vary between 20,000-50,000 considering a uniform isotropic background  UV flux and are shown in Fig. \ref{figure4}. Similar values of $J_{21}^{\rm crit}$ have been obtained for anisotropic source \citep{Regan2014B,Regan2015b}. Moreover, the presence of X-rays may further influence the $J_{21}^{\rm crit}$  depending on its strength \citep{Latif2015a,Inayoshi2015,Regan2016} and also an accurate modelling of $\rm H_2$ self-shielding is required \citep{WolocottGreen2011,Hartwig2015}. 

\subsection{Metal pollution}
One of the key requirements for the formation of DCBHs via isothermal collapse is that the hosting halo should be metal free.  The metal enrichment  is expected to be patchy in the early universe and therefore some halos  may remain unpolluted even down to z= 6 \citep{Tornatore2007,Trenti2009,Maio2011,Ritter2014,Pallottini2014,Habouzit2016}. In fact, observations show that pockets of extremely metal poor gas ($\rm Z/Z_{\odot} \sim 10^{-5}$) can exist down to z=7 \citep{Simcoe2012}.  So, it is conceivable that primordial halos where star formation is suppressed in the presence of LW background may stay metal free at $z \geq10$. 

 However, the DCBH host halo is expected to form in the surrounding of a star forming galaxy to receive a strong LW flux and therefore has to avoid the possible pollution by the supernova winds. \cite{Dijksta2014} employed an analytical model to study the impact of metal pollution by the supernova winds and found that it can significantly affect the expected abundance of DCBHs. \cite{Visbal2014} propose that  metal pollution  can also be avoided in a synchronised pair of halos.  In such a case, star forming halo forms first while  the DCBH host halo forms later and rapidly collapses before it gets enriched by the supernova winds.  Moreover, such pair of halos is expected to be in a clustered environment where metals might be ejected in a preferential direction with  low density and  metal pollution may be avoided \citep{Ritter2014,Pallottini2014}. It has been recently found that some of the DCBH host halos may get tidally disrupted in such a scenario \citep{Chon2016}. 
 
\begin{figure}
\hspace{-0.2cm}
\includegraphics[scale=0.45]{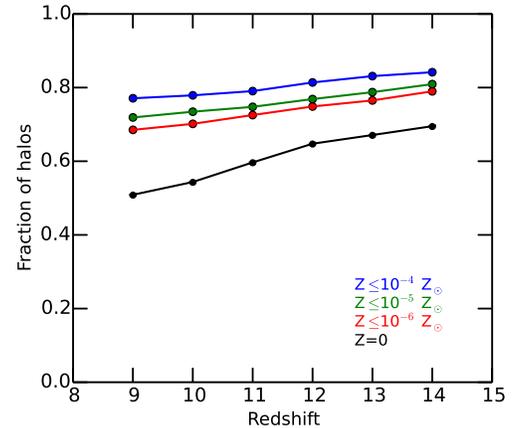} 
\caption{ Fraction of halos with metallicity below the given value in the figure legged and masses between $\rm 2 \times ~10^7-10^8~M_{\odot} $.  Adopted from \cite{Latif2015Dust}. }
\label{figure5}
\end{figure}

Cosmological hydrodynamical simulations including both star formation and supernova feedback show that a significant fraction of halos remains metal free down to z=10  with mass range  between $\rm 2 \times 10^7- 10^8~M_{\odot}$ \citep{Latif2015Dust,Habouzit2016}, see Fig. \ref{figure5}. These results provide an upper limit on the fraction of metal free halos  in the above mentioned range as simulations are unable to resolve halos below $\rm 10^7 ~M_{\odot}$. In future,  cosmological hydrodynamical simulations self-consistently taking into account the metal pollution as well as radiations from the star forming galaxies  are required to better understand the metal pollution in the DCBH host halos.

\subsection{Formation of a supermassive/quasi star}
Supermassive stars (SMSs)\footnote{SMS is ordinary star with mass above 1000 $\rm M_{\odot}$ and quasi-star is a branch of SMS whose core collapses into a BH, see \cite{Schleicher13} for a detailed discussion} are considered as potential cradles for the formation of DCBHs \citep{Begelman2010,Volonteri2010b,Ball2011,Hosokawa12,Hosokawa2013,Schleicher13,Johnson2013b,Ferrara14}. They may collapse via general relativistic (GR) instabilities  into a  DCBH while retaining $\rm \geq 50\%$ of their initial mass \citep{Baumgarte1999,Shibata2002,Montero2012,Reisswig2013} or may evolve towards a ZAMS and later  collapse into a massive BH \citep{Ferrara14}.  The stellar evolution calculations suggest that the formation of such objects requires rapid accretion with $\rm \dot{m} \geq 0.1~ M_{\odot}/yr$ \citep{Begelman2010,Ball2011,Hosokawa2013,Schleicher13}. For such high accretion rates, the radius of star monotonically increases with mass  due to the shorter accretion time in comparison with the Kelvin-Helmholtz contraction time scale \citep{Hosokawa2013, Sakurai2015}.  Consequently,  their surface temperatures remain as low as 5000 K, they  produce  weak UV  stellar feedback and can grow up to $\rm 3.6 \times 10^8 ~ \dot{m} ~M_{\odot}$. The energy released by the nuclear burning remains subdominant compared to the energy produced by the stellar contraction.

\cite{Schleicher13} found that for $\rm \dot{m} \geq 0.14~ M_{\odot}/yr$ the core of a SMS collapses into a BH and forms a so-called quasi-star, also see \cite{Begelman2008}. It comes from the fact that accretion time scale is considerably shorter than the nuclear burning time scale and therefore the core collapses into a black hole \citep{Begelman2006,Begelman2010}. A possible advantage of such composition is that a central BH may accrete at the super-Eddington rate but overall accretion is limited by the Eddington rate  of a quasi-star. \cite{Ball2011} have found that about 10 \% of the stellar mass is accreted onto the BH before the hydrostatic equilibrium breaks down and results are sensitive to the choice of boundary conditions. \cite{Dotan2011} quantified the potential impact of radiation driven winds from the envelope of a quasi-star and  found that winds can be so strong that they may blow away the envelope before the BH doubles its mass. The effect of winds from the radiation dominated objects was reconsidered by \cite{Fiacconi2016} by solving the equations of motion and including the previously neglected advection energy term.  They found that the super Eddington accretion onto  the newly born BH within quasi-stars is likely responsible for vigorous mass loss and limits the BH growth. In the follow up study \cite{Fiacconi2016a} explored the impact of rotation in quasi-star scenario via simplified analytical model and found that for  $\rm  \geq 10^5 ~M_{\odot}$ massive SMSs the quasi-star phase may be skipped while the growth of less massive objects may get prevented. However, 3D radiation hydrodynamics simulations are necessary to  validate these findings.

\begin{figure*}
\hspace{0.1cm}
\includegraphics[scale=0.2]{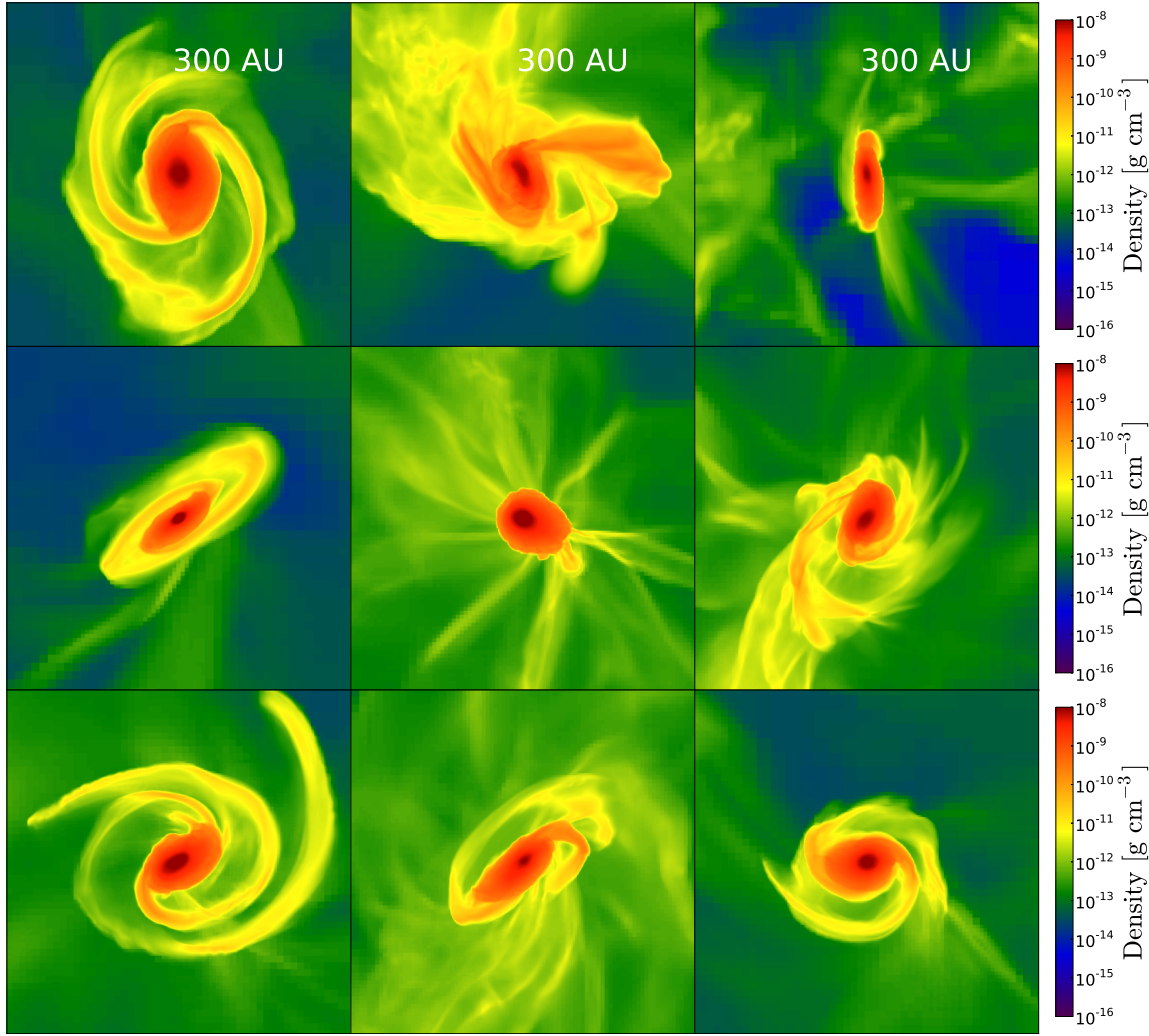} 
\caption{ Self-gravitating accretion disks formed at the centre of a massive primordial halos illuminated by a strong LW flux as a consequence of isothermal collapse. Each panel represents a halo of above $\rm 10^7~M_{\odot}$ forming at $z=10-15$  and shows the density projection in the central 300 AU. Adopted from \cite{Latif2013c}.  }
\label{figure6}
\end{figure*}

The numerical experiments employing a Jeans resolution of  4 cells did not observe any fragmentation \citep{Bromm03,Wise2008,Regan09,Prieto2013,Latif2013}  while the simulations employing both higher Jeans  (64 cells per Jeans length) as well as spatial resolution show  that fragmentation occasionally occurs \citep{Latif2013c, Latif2013d,Regan2014a, Bcerra2014}. They further found that  a self-gravitating accretion disk forms in the centre of the halo (see Fig. \ref{figure6}) which becomes marginally unstable and fragments into multiple clumps but most of the clumps get merged with the central clump. Large accretion rates of $\rm 0.1-1~M_{\odot}/yr$ are observed in these simulations and seem to be sufficient  for forming a massive central object. Moreover, subgrid scale turbulence and strong magnetic fields amplified via the small scale dynamo help in suppressing  fragmentation, see \cite{Latif2013c} and  \cite{Latif2013a,LatifMag2014}. These simulations did not include $\rm H^-$ cooling which becomes important at densities  of $\rm 10^{8}-10^{16} ~cm^{-3}$ and also cooling was artificially switched off  to mimic the formation of a protostar.  Both idealised \citep{VanBorm2014,Inayoshi2014} and cosmological \citep{Latif2016} simulations employing  a detailed chemical model have been performed. Particularly, they included the $\rm H^-$ cooling and opacities for  $\rm H^-$ bound-free emission, free-free absorption as well as the Raleigh scattering of hydrogen atoms. These simulations show that $\rm H^-$ cooling does not halt the formation of a SMS and may lead to a binary formation in some cases but realistic opacities help in  stabilising the collapse on small scales \citep{Latif2016}.

Cosmological simulations using sink particles (also without sinks) confirm the formation of a SMS of $\rm 10^5~M_{\odot}$ within a 1 Myr after its birth \citep{Latif2013d,Shlosman2016}. Moreover, stellar evolution simulations employing both constant  ($\rm \geq 0.1 ~M_{\odot}/yr$) and time dependent mass accretion rates  show that  a SMS can be formed and resulting UV feedback is too weak to hinder the mass accretion \citep{Hosokawa2013,Sakurai2015,Sakurai2016}. The present studies suggest the formation of a SMS as a potential outcome \citep{Hosokawa2013,Sakurai2016} but do not consider the effect of rotation. However,  it is not yet clear if the final outcome of an isothermal  collapse is always a SMS or a quasi-star or even a massive BH. In future 3D cosmological hydrodynamical simulations including the UV feedback from a  SMS and coupled stellar evolution will be required to asses it.
 
\subsection{Birth mass function}
 The studies investigating the formation of a SMS could not evolve  their simulations long enough to assess its final fate due to the numerical constraints \citep{Hosokawa2012, Hosokawa2013}. So, it is not clear whether a SMS directly collapses into a BH via GR instabilities provided that accretion continues  with $\rm \geq 0.1~M_{\odot}/yr$ or it evolves to a ZAMS (if the mass accretion onto the SMS stops ) and  later collapses into a massive BH.  It is also not well understood what is the birth mass function of DCBHs.  Moreover, previous works ignored the effect of rotation on the evolution of a SMS which can play important role in determining its final fate. \cite{Ferrara14} used an analytical model to investigate the fate of an accreting  SMS and found that it becomes GR unstable when its equation of state drops below a critical value \citep{Chandra1964,Montero2012} while the presence of rotation  halts the collapse.  The  mass  of a SMS  can be estimated as \citep{Ferrara14}:
\begin{equation}
\mathrm{M_{*}} \leq  8.48 \times 10^{5}  \left(  \frac{\dot{m}}{M_{\odot} yr^{-1}} \right)^{2/3} ~~~ No ~Rotation \\ 
\label{r1}
\end{equation}

\begin{equation}
\mathrm{M_{*}} \leq 6.01 \times 10^{5}  \left(  \frac{\dot{m}}{M_{\odot} yr^{-1}} \right) ~~~~~~~~~~~  Rotation \\ 
\label{r2}
\end{equation}
SMSs more massive than these limits are expected to directly collapse into a BH.

\begin{figure}
%\hspace{-0.0cm}
\includegraphics[scale=1.0]{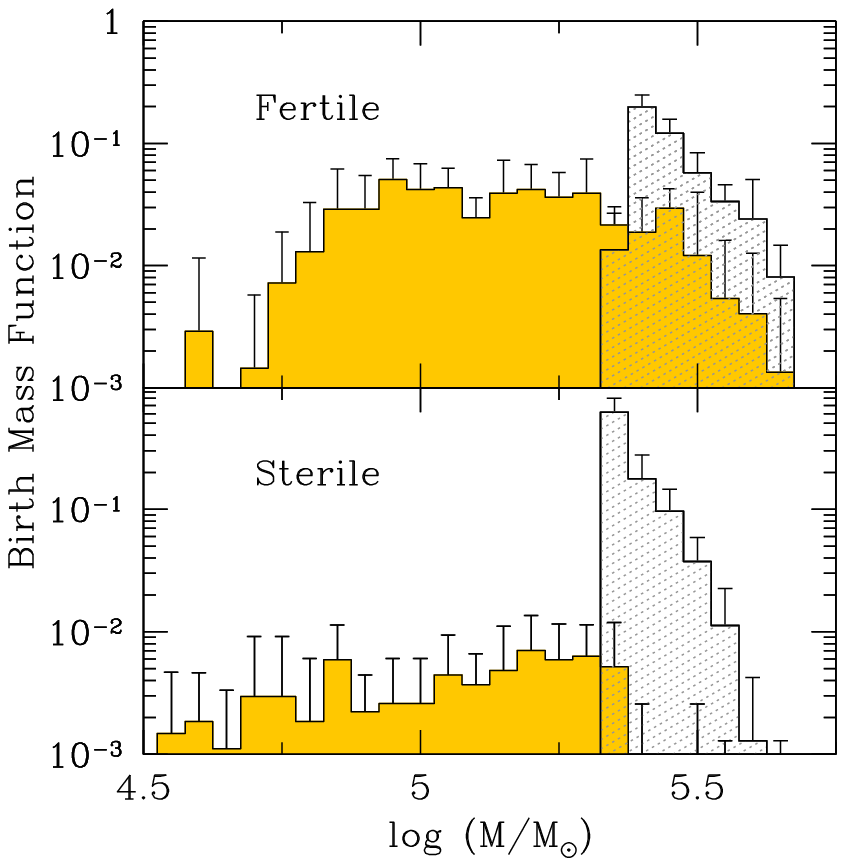} 
\caption{ The mass distribution of  DCBH seeds (dotted histogram) and SMS (yellow histogram). The upper panel shows the case of fertile minihalos while the bottom panel sterile minihalos. The results are computed from merger tree simulations and averaged over 50 milky-way merger histories with $\pm \sigma$ error bars. Adopted from \cite{Ferrara14}. }
\label{figure7}
\end{figure}

\cite{Ferrara14} used merger tree  simulations to compute the birth mass function of DCBHs  by following the growth of a SMS. Depending on the mass accretion onto the proto-SMS they determined whether it forms a DCBH via GR instabilities or  a protostar contracts and evolves into a ZAMS SMS. The mass distribution of DCBHs is shown in Fig. \ref{figure7} and depicts that  typical masses of DCBHs are $\rm \sim 2 \times 10^5 ~M_{\odot}$ while the masses of SMSs  vary from $\rm  2 \times 10^4- 10^5 ~M_{\odot}$. The  final mass further depends on whether the merged minihalo/accreted gas was metal free (dubbed as the sterile case) or metal enriched (fertile case). Overall, the masses of DCBHs are in the range of $\rm 10^{4-5}~M_{\odot}$.

%\section{Primordial BHs}

\section{Future Outlook}
In this review we mainly focused on the formation mechanisms of supermassive BH seeds and other aspects such as their number density and growth are covered in the complementary reviews. However, for the sake of completeness, we briefly mention them here. The number density of seed BHs depends on their formation mechanism, see \cite{Agarwal2012},\cite{Dijksta2014}, and \cite{Habouzit2016a}.  Due to the special conditions required for the formation of heavy seeds their number density is expected to be smaller compared to the lighter seeds. Depending on the seed BH mass ($\rm 10-10^5~M_{\odot}$), they require 10-20 e-foldings to reach billion solar masses  at $z \geq 6$, see \cite{Johnson2016PASA} for a detailed discussion on the early growth of seed BHs.  This may be achieved via prolonged episodes of accretion and/or merging in rare massive halos forming at earlier cosmic times  as found in large  cosmological simulations \citep{Sijacki2007,Sijacki2015,DiMatteo2012,DiMatteo2016}.

Although tremendous progress has been made during the past decade both from  the theoretical and the observational perspectives to understand the formation mechanisms of supermassive black holes, there are still many open questions. Theorists have proposed several models but direct observational evidence is required to constrain them. This might have been provided for the first time by the recent discovery claim made by \cite{Pacucci2016}. This work has suggested a novel method to identify BH seeds in deep galaxy surveys that, in addition to detecting two such sources in the currently available data, seems optimal also for future JWST searches.  To conclude, we briefly summarize in the following some of the most urgent questions concerning the formation mechanisms of seed black holes:
\begin{itemize}
\item How massive are the first stars and what is their IMF?
\item How compact and massive are the first nuclear star clusters?
\item What is the binary fraction of Pop III and Pop II stars? 
\item Is the formation of  a VMS  in a dense stellar cluster viable and what is the typical mass of  the resulting black hole?
\item What are the intermediate stages in the direct collapse BH scenario? a SMS, a quasi-star or none of these?
\item How abundant are black seeds in various formation mechanisms?
\item Under what conditions seed BHs can grow to the SMBHs scales of a few billions solar masses?
\item How can we observationally constrain different BH formation models?
\end{itemize}

Several ways have been suggested to  observationally constrain the BH formation models.  One of these is to probe the low luminosity active galactic nuclei (AGNs) at $z>6$ and measure their mass accretion accretion rates and BH masses. In fact, ATHENA X-ray observatory is expected to detect about 400  low luminosity AGNs with $\rm L_{X} \geq 10^{43}~erg/s$ at $z\geq6$ and will provide  direct constraints on BH formation mechanisms \citep{Aird2013}.  The low yields of AGNs at $z>6$ will rule out the PopIII seed formation scenario as they are expected to be more abundant compared to the heavy seeds, see Fig. 1 and discussion in  \cite{Aird2013}.  For example,  an AGN with  $\rm L_{X} = 4 \times 10^{43}~erg/s$  observed at $z\geq 10$ needs to be powered by a  SMBH of $\rm > 10^7~M_{\odot}$. For an Eddington limited growth within the age of universe at $z =10$, this requires a seed BH of $\rm 10^5~M_{\odot}$. Such observations can directly constrain the masses, growth rates and duty cycles of seed BHs.

The recent detection of gravitational waves from the merging  of a BH binary by LIGO \citep{Abbott2016} has opened the possibility to constrain the population of stellar mass BHs in the early Universe. Depending on the IMF, Pop III BH binaries may produce distinct features in the gravitational wave background in 10-100 Hz band  and yield constraints on the abundance of  Pop III BHs \citep{Hartwig2016b,Inayoshi2016c,Ricotti2016}. Alternative strategies  rely on secondary indicators such as measuring the occupation fraction of  MBHs  in low mass galaxies or constraints from growth time scales of BHs \citep{Volonteri2010,Greene2012}.  In fact, recent observations of low mass galaxies indicate the potential presence of a $\rm 10^5 -10^6~M_{\odot}$ BHs at their centers \citep{Reines2011,Satyabal2014}. Confirmation of such AGNs with future observations and precise measurements of their BH masses will tremendously help to distinguish the BH formation models.

Recent detection of the brightest Lyman alpha emitter at $z=6.6$  (CR7) by \cite{Sobral2015} shows strong Lyman alpha and He-1640$\AA$  line luminosities and no metal line emission. Numerous studies suggest that CR7 may potentially host a DCBH \citep{Pallottini2015,Hartwig2015b,Agarwal2015c,Dijkstra2016,Smidt2016,Smith2016}. However, further observations are required to confirm this hypothesis.  Upcoming space and ground based telescopes such as JWST,  ATHENA, WFIRST and SKA  will be able to directly probe seed BHs in the high redshift universe and provide  direct observational constraints.

\begin{acknowledgements}
The authors  gratefully acknowledge R. Schneider, R. Valiante and M. Volonteri for stimulating this review. This project has received funding from the European Union's Horizon 2020 research and innovation programme under the Marie Sklodowska-Curie grant  agreement N$^o$ 656428. This research was supported in part by the National Science Foundation under Grant No. NSF PHY11-25915.
\end{acknowledgements}

%\nocite*{}
\bibliographystyle{pasa-mnras}
\bibliography{smbhs.bib}
\end{document}